\documentclass[preprintnumbers,amsmath,amssymb,12pt,floatfix,epsfig,A4,superscriptaddress,preprint]{revtex4}
\usepackage{bm}
\usepackage{natbib}
\usepackage[dvips]{color}
\usepackage{url}
\usepackage{graphicx}
\usepackage{amsmath}
\usepackage{float}
\usepackage{xcolor}
\usepackage{ulem}
\date{\today}

\usepackage{color} % to paint the words to be indexed

\begin{document}
\title{Folding potential with modern nuclear density functionals and application to $^{16}$O+$^{208}$Pb reaction}

\author{Kyoungsu Heo}
\affiliation{Department of Physics and Origin of Matter and Evolution of Galaxies Institute, Soongsil University, Seoul 06978, Korea}
\author{Hana Gil}
\address{Center for Extreme Nuclear Matters, Korea University, Seoul 02841, Korea}
\author{Ki-Seok Choi}
\address{School of Liberal Arts and Science, Korea Aerospace University, Koyang 10540, Korea}
\author{K. S. Kim}
\address{School of Liberal Arts and Science, Korea Aerospace University, Koyang 10540, Korea}
\author{Chang Ho Hyun}
\address{Department of Physics Education, Daegu University, Gyeongsan 38453, Korea}
\author{W. Y. So}
\thanks{\textrm{e-mail:} wyso@kangwon.ac.kr}
\address{Department of Radiological Science, Kangwon National University at Dogye, Samcheok 25949, Korea}

\date{\today}
\begin{abstract}
Double folding potential is constructed using the M3Y interaction and the matter densities of the projectile and target nuclei obtained from four microscopic energy density functional (EDF) models. The elastic scattering cross sections for the $^{16}$O+$^{208}$Pb system are calculated using the optical model with the double folding potentials of the four EDF models. We focus on the correlation between the matter densities and the behavior the double folding potential and the elastic scattering cross sections. First, the matter and charge densities are examined by comparing the results of the four EDF models. There is a slight difference in the density in the internal region, but it is negligible in the outer region. Next, we calculate the double folding potential with
the matter densities obtained from the four EDF models. Differences between the models are negligible in the outer region, but the potential depth in the internal region shows model dependence, which can be understood from the behavior of matter densities in the internal region. Another point is that the double folding potential is shown to be weakly dependent on the incident energy. Finally,  the elastic scattering cross sections have no significant model dependence except for the slight difference in the backward angle.

\end{abstract}

\pacs{24.10.-i, 25.70.Jj}

%\keywords{Double folding potential, Optical model, Microscopic energy density functional model}

\maketitle

\section{Introduction}

Nuclear reactions are the cornerstones to understand the origin of heavy elements which is one of the most
fundamental questions in nature.
Particularly, the importance of the reactions is in the supernova explosion and neutron star merger,
which are believed to be the major factory for the creation of stable and unstable isotopes.
According to numerical simulations, heavy elements are synthesized along the neutron drip line at first,
and their unstable isotopes undergo various decay and reaction processes, and finally arrive at the stability valley.
Understanding the reactions of these unstable nuclei may provide a clue of opening the secret of the origin of heavy elements and the evolution of Universe.

An essential difficulty with the reactions of the neutron-rich nuclei is that it is hard to produce these rare isotopes
in the laboratory experiment, and it is even harder to achieve desired reactions with the nuclei close to the neutron drip line.
On the other hand, stable nuclei allow easy access to experiments and accurate  measurements of the reaction rates.
For this reason, theory of the nuclear reaction has been developed based on the reactions of stable nuclei.
Theoretical schemes for the reaction processes have been successful in reproducing the experimental data with only
a few parameters.

Folding potential is one of the successful methods for a simple but accurate description of the nuclear reaction.
Double folding potential is composed two main ingredients, one is the nucleon-nucleon ($NN$) interaction,
and the other is the density profile of the target and projectile nuclei.
Michigan-3-Yukawa (M3Y) potential \cite{sat1}
 has long been a representative model for  the $NN$ part of the double folding potential.
For the density profile of the nuclei, phenomenological models based on the Woods-Saxon density function has been
a principal way for the description of the nuclei.
With only two or three parameters in the Woods-Saxon density function,
the method provides accurate description of the basic nuclear properties like the binding energy, charge radius and matter distribution.
However, the parameters in the density function must be fitted to the well-known properties of each nucleus,
so that it becomes difficult to determine the parameters in the Woods-Saxon density function without any exact information of a specific nucleus.
Therefore, if the method is to be applied to the reactions of the nuclei close to the neutron drip line, it confronts a serious limitation.

Main focus of the present work is to improve the nuclear density part of the folding potential by replacing the Woods-Saxon density function
with the wave functions obtained from microscopic models based on the nuclear density functional theory.
In the nuclear energy density functional (EDF) formalism, single-particle wave functions are obtained by solving the Hartree-Fock or Dirac equations
with potentials fitted to nuclear data.
Wave functions thus obtained reproduce accurately not only the bulk properties of nuclei,
but also the internal structures such as single-particle energy levels, quadrupole deformation and etc.
In a recent work \cite{prc2023}, predictive power of the formalism has been shown by calculating
the properties even-even nuclei across the nuclear chart with parameters constrained by only a few nuclear data.
Though only six nuclear data have been used in fitting the model parameters,
the result shows agreement with nuclear data at high accuracy,
so it is demonstrated that the density functional theory provides a framework suitable and reliable for extension to the reactions
of the nuclei whose properties are not known yet.
The models in Ref.~\cite{prc2023} were also applied to the lepton-nucleus scattering \cite{lepton1,lepton2},
and the results are in good agreement with the experimental data.

In this work, we consider four EDF models, two from the relativistic mean field theory models [quantum hadrodynamics (QHD) and quark meson coupling (QMC)]
and the other two from the nonrelativistic Skyrme force models [SLy4 and Korea-IBS-Daegu-SKKU (KIDS)].
As a first step for the application of the EDFs to the nuclear reactions, we consider the well-known
reactions of the  $^{16}$O+$^{208}$Pb system at various incident energies.
In the result, we first examine the charge and matter distributions of $^{16}$O and $^{208}$Pb.
Charge distributions are compared with the data.
In the next step, we calculate the double folding potentials of the $^{16}$O+$^{208}$Pb system
at $E_{\textrm{lab}}$ = 80, 90, and 192 MeV corresponding to below, near, and above barrier energies.
It is seen that the folding potential is weakly dependent on the energy,
and its behavior is directly related to the matter distribution of the target and projectile nuclei.
In order to calculate the elastic scattering cross section, the imaginary part of the optical model potential must be fitted to data.
In general fitting is done for each model at each incident energy.
In this work, instead of conventional approach, we determine the imaginary potential with a specific model
and apply the potential to other models.
This way can give us an idea on the sensitivity of the imaginary potential to a specific model.
The result shows that the dependence of the imaginary potential on a model is weak,
and the cross sections calculated with the four models agree well with the scattering data at $E_{\textrm{lab}}$ = 80, 90 and 192 MeV.
However at certain kinematics, model dependence appears obvious.
This observation opens intriguing questions whether such a dependence will happen in the reactions of neutron-rich rare isotopes,
and how much it will affect the reaction rates of $r$-process and consequently the abundance of heavy elements in Universe.

In the following parts of the article, we briefly introduce the models employed in the work,
present numerical results and discussions on them, and finally summarize the work.

%%%%%%%%%%%%%%%%%%
\section{Nuclear models in the folding potential}
%%%%%%%%%%%%%%%%%%

Double folding potential $V_{\rm DF}(r)$ which describes the interaction between target and projectile nuclei can be written as~\cite{sat1}
\begin{equation}
\label{fold_pot}
V_{\textrm{DF}}({\bf r})=\int d{\bf r}_{1} d{\bf r}_{2} \rho_{1}({\bf r}_{1})
\rho_{2}({\bf r}_{2}) v_{NN}(r_{12}),
\end{equation}
where $\rho_{1}({\bf r}_{1})$ and $\rho_{2}({\bf r}_{2})$ are the nuclear matter distributions within the projectile and target nuclei, respectively
and $r_{12} =|\bf{r}-\bf{r}_{1}+\bf{r}_{2}|$.
$v_{NN}$ denotes the nucleon-nucleon ($NN$) interaction.
Most conventional $NN$ interaction employed in the folding potential is the M3Y model given by
\begin{equation}
\label{v_nn}
v_{NN}(r) = 7999\frac{e^{-4r}}{4r} - 2134\frac{e^{-2.5r}}{2.5r} - 276 \Big(1-0.005\frac{E_{\textrm{lab}}}{A_{1}} \Big)\delta (r).
\end{equation}
The first two terms corresponding to the direct part are composed of two Yukawa functions,
and the last term representing the exchange part has a delta-function form.
$E_{\textrm{lab}}$ and $A_{1}$ are the incident laboratory energy and the mass number for the projectile, respectively.

Present work focuses on the effect of the density profile $\rho(\mathbf{r})$ obtained from microscopic nuclear models.
We consider four models, two from the nonrelativistic Skyrme force models and two from the relativistic mean field theory.
We briefly describe the models in the following subsections.

\subsection{Skyrme force: SLy4 and KIDS}

In the nonrelativistic microscopic models,
the wave function of a nucleon in nuclei is obtained by  solving the Hartree-Fock equation
\begin{equation}
\left[ {\bf \nabla} \frac{\hbar^2}{2 m^*_q({\bf r})} \cdot {\bf \nabla}
+ U_q({\bf r}) - i {\bf W}_q \cdot ({\bf \nabla} \times \mbox{\boldmath$\sigma$}) \right] \psi_i ({\bf r}, q) = E_i \psi_i ({\bf r}, q),
\end{equation}
where $U_q$ and ${\bf W}_q$ are the mean field single-particle potentials for the central and spin-orbit interactions, respectively.
Effective mass of the nucleon is given by $m^*_q({\bf r})$ where $q$ denotes the proton or neutron,
and {\boldmath$\sigma$} is the spin operator.
A most traditional approach is to start with an effective potential that contains two- and three-body nuclear interactions,
and calculate the matrix elements of the effective potential to obtain $U_q$ and ${\bf W}_q$.
In this work, we adopt the SLy4 Skyrme force \cite{sly4} which is one of the most popular effective potential models
in the nonrelativistic approach.
In the conventional Skyrme force model, $m^*_q({\bm r})$, $U_q$ and ${\bf W}_q$ are given by
\begin{eqnarray}
\frac{\hbar^2}{2m^*_q({\bf r})} &=& \frac{\hbar^2}{2 m_q} + \frac{1}{8} [t_1 (2+x_1) + t_2 (2+x_2)] \rho({\bf r})
+ \frac{1}{8} [-t_1 (1+2x_1) + t_2 (1+2 x_2)] \rho_q({\bf r})
\label{eq:mstar}
\end{eqnarray}
where $m_q$ denotes the nucleon mass in free space, and
\begin{eqnarray}
U_q({\bf r}) &=& \frac{1}{2} t_0 [ (2+x_0) \rho - (1+2x_0) \rho_q] \nonumber \\
&+& \frac{1}{24} t_3 \{(2+x_3)(2+\alpha) \rho^{\alpha+1}
- (2x_3+1)[2\rho^\alpha \rho_q + \alpha \rho^{\alpha-1}(\rho^2_p + \rho^2_n)\} \nonumber \\
&+& \frac{1}{8} [t_1(2+x_1)+t_2(2+x_2)]\tau + \frac{1}{8} [t_2(2x_2+1) - t_1(2x_1+1)] \tau_q \nonumber \\
&+& \frac{1}{16} [t_2(2+x_2)-t_1(2+x_1)] \nabla^2 \rho + \frac{1}{16}[3t_1(2x_1+1) + t_2(2x_2+1)]\nabla^2 \rho_q \nonumber\\
&+& \frac{1}{8}(t_1-t_2) {\bf J}_q - \frac{1}{8} (t_1 x_1 + t_2 x_2) {\bf J},
\label{eq:uq}
\end{eqnarray}
where $\tau_q$ and ${\bf J}_q$ are the kinetic and spin densities, respectively, and
\begin{eqnarray}
{\bf W}_q({\bf r}) = \frac{1}{2} W_0 (\nabla \rho + \nabla \rho_q).
\label{eq:wq}
\end{eqnarray}
Numerical values of the parameters $t_0$, $t_1$, $t_2$, $t_3$, $x_0$, $x_1$, $x_2$, $x_3$, $\alpha$, and $W_0$
for the SLy4 model can be found in Ref.~\cite{sly4}.

Another approach is the density functional theory.
In the KIDS functional framework,
assuming that the energy of a multi-nucleon system can be written down uniquely as a function of the density of the proton and neutron,
energy density of infinite nuclear matter and finite nuclei is expanded as a series of $\rho^{1/3}$ \cite{kids-nm}.
Unknown parameters of the functional are adjusted to the data of infinite nuclear matter and finite nuclei~\cite{kids-acta,prc99}.
Mean field potential of the KIDS functional can be obtained by transforming the interaction terms in the energy density
to the Skyrme-type contact force \cite{npsm2017, symmetry2023}.
In the conventional Skyrme force, three-body interaction is described in terms of a density-dependent term ($\alpha$ in Eq.~(\ref{eq:uq})
is the power of the density).
An essential difference between the normal Skyrme force and the KIDS formalism is the multiple density dependence.
In the KIDS framework, the term proportional to $t_3$ in Eq. (\ref{eq:uq}) is replaced with
\begin{eqnarray}
\frac{1}{24} \sum^3_{k=1} \rho^{k/3} \left[ \left( 2 + \frac{k}{3}\right) \left(2t_{3k} + y_{3k}\right) \rho
- 2 (t_{3k} + 2 y_{3k}) \rho_q - \frac{k}{3} \left(t_{3k} + 2 y_{3k} \right) \frac{\rho^2_p + \rho^2_n}{\rho} \right],
\end{eqnarray}
and the other terms in $U_q({\bf r})$, and $m^*_q({\bf r})$ and ${\bf W}_q({\bf r})$ take the same form
as given in Eqs. (\ref{eq:mstar}, \ref{eq:uq}, \ref{eq:wq}).
The potential thus obtained are directly adoptable in the Skyrme-Hartree-Fock codes.

\subsection{Relativistic mean fields: QHD and QMC}

The effective Lagrangian density with the $\sigma$, $\omega$ and $\rho$ meson field \cite{npa1981,qmc1} in the mean field approximation is written for static spherical finite nuclei:
\begin{eqnarray}
{\cal L} &=& {\bar \psi} \left [ i \gamma \cdot \partial -M^*_N (\sigma ({\bf r})) - g_{\omega} \omega ({\bf r}) \gamma_0 -g_{\rho} {\frac {\tau_3^N} 2} b({\bf r}) \gamma^0 - {\frac e 2} (1 + \tau^N_3 ) A({\bf r}) \gamma^0 \right ] \psi \nonumber \\
&-&{\frac 1 2} \left [ ( \nabla \sigma ({\bf r}))^2 + m_{\sigma}^2 {\sigma ({\bf r})}^2 \right ] + {\frac 1 2} \left [ ( \nabla \omega ({\mathbf{r}}))^2 + m_{\omega}^2 {\omega ({\mathbf{r}})}^2 \right ] \nonumber \\
&+& {\frac 1 2} \left [ ( \nabla b ({\bf r}))^2 + m_{\rho}^2 {b ({\bf r})}^2 \right ] + {\frac 1 2} [ \nabla A ({\bf r}) ]^2, \label{lagran}
\end{eqnarray}
where $\psi ({\bf r})$, $\sigma ({\bf r})$, $\omega ({\bf r})$ and $b({\bf r})$ are the fields of the nucleon, $\sigma$ meson, time component of the $\omega$ meson and time and isospin third component of the $\rho$ meson, respectively.
$A({\bf r})$ is the time component of the Coulomb field.
$m_\sigma$, $m_\omega$ and $m_\rho$ are the masses of the $\sigma$, $\omega$ and $\rho$ mesons, respectively.
Vector meson-nucleon coupling constants are written in terms of $g_\omega$ and $g_\rho$.
$\tau^N_3 /2$ is the third component of the nucleon isospin operator.
$M^*_{N}$ denotes the effective nucleon mass and is defined as
\begin{equation}
M^*_{N} ({\sigma ({\bf r})}) = M_N - g_{\sigma} ({\sigma ({\bf r})}) \sigma ({\bf r}),
\end{equation}
with the free nucleon mass $M_N$.
The function $g_{\sigma} ({\sigma ({\bf r})})$ is given by
\begin{equation}
g_{\sigma} ({\sigma ({\bf r})}) = g_{\sigma} \left [ 1 - {\frac a 2} g_{\sigma} \sigma ({\bf r}) \right ]
\end{equation}
with a parameter $a$ related to the bag radius~\cite{qmc1}.
QHD model treats the nucleon a point particle, so $a = 0$, which leads to a simple relation for the $\sigma-N$ coupling constant $g_\sigma (\sigma({\bf r})) = g_\sigma$.
Since the nucleon is treated as a composite system in the QMC model, $a \neq 0$, so explicit form of $g_\sigma(\sigma({\bf r}))$ is more complicated.
The detailed description of the QMC model is in Refs. \cite{qmc1,qmc2}.

From the effective Lagrangian density in Eq.~(\ref{lagran}), the Dirac equation is written as
\begin{equation}
\left [ \gamma_{\mu} p^{\mu}-M-S(r)-\gamma^{0}V(r)\right  ] \Psi({\bf r}) = 0
\label{dirac}
\end{equation}
with
\begin{eqnarray}\label{dirac-wave}
{\Psi}({\bf r})=\left( \begin{array}{c} f_{\kappa}(r){\chi}^{{\mu}}_{{\kappa}}({\hat {\bf r}})\\
                                        i g_{\kappa}(r){\chi}^{{\mu}}_{-{\kappa}}({\hat {\bf r}}) \end{array} \right)  \;.
\end{eqnarray}
Here $S(r)$ and $V(r)$ are the symmetric scalar potential and time-component vector potential, respectively.
Single-particle wave function ${\Psi}({\bf r})$ is made up of the product of the radial parts and the spin parts.
The spin function is given by
\begin{equation}
\chi^{\mu}_{\kappa} ({\hat {\bf r}}) = \sum_{m,s} \langle l m {\frac 1 2} s | j \mu \rangle Y_{l,m} ( {\hat {\bf r}}) \chi_s
\end{equation}
with the Pauli spinor $\chi_s$ and the spherical harmonics $Y_{l,m}$.
$\kappa$ in Eq.~(\ref{dirac-wave}) is the Dirac quantum number. As a result, the radial parts of the Dirac equation are obtained from Eqs.~(\ref {dirac}) and~(\ref {dirac-wave}) as follow:
\begin{eqnarray}
{\frac {df_{\kappa}} {dr}} &=& - {\frac {\kappa +1} {r}} f_{\kappa} +
(M+E+S-V) g_{\kappa} \;, \nonumber \\
{\frac {dg_{\kappa}} {dr}} &=& {\frac {\kappa -1} {r}} g_{\kappa} +
(M-E+S+V) f_{\kappa} \;,
\label{radeq2}
\end{eqnarray}
where $E$ is the energy eigenvalue of a nucleon in nuclei.
Please see Refs.~\cite{npa1981,kim1} for details.

%%%%%%%%%%%%%%%%%%%%%%%%%%%%
\section{Results}

\subsection{Density profile of $^{16}$O and $^{208}$Pb}

%%% Figure 1 %%%
\begin{figure}
\begin{center}
\includegraphics[width=0.495\linewidth] {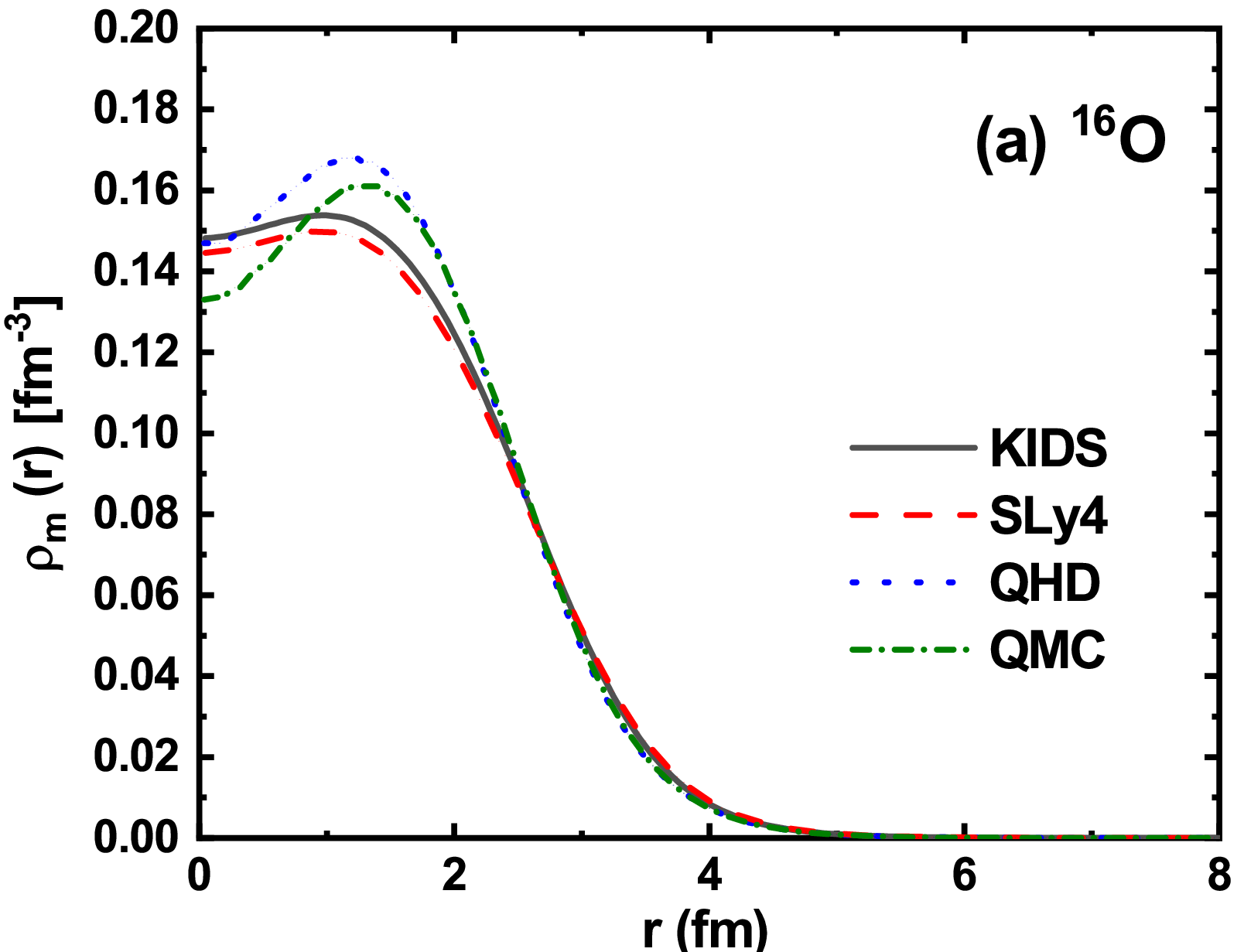}
\includegraphics[width=0.495\linewidth] {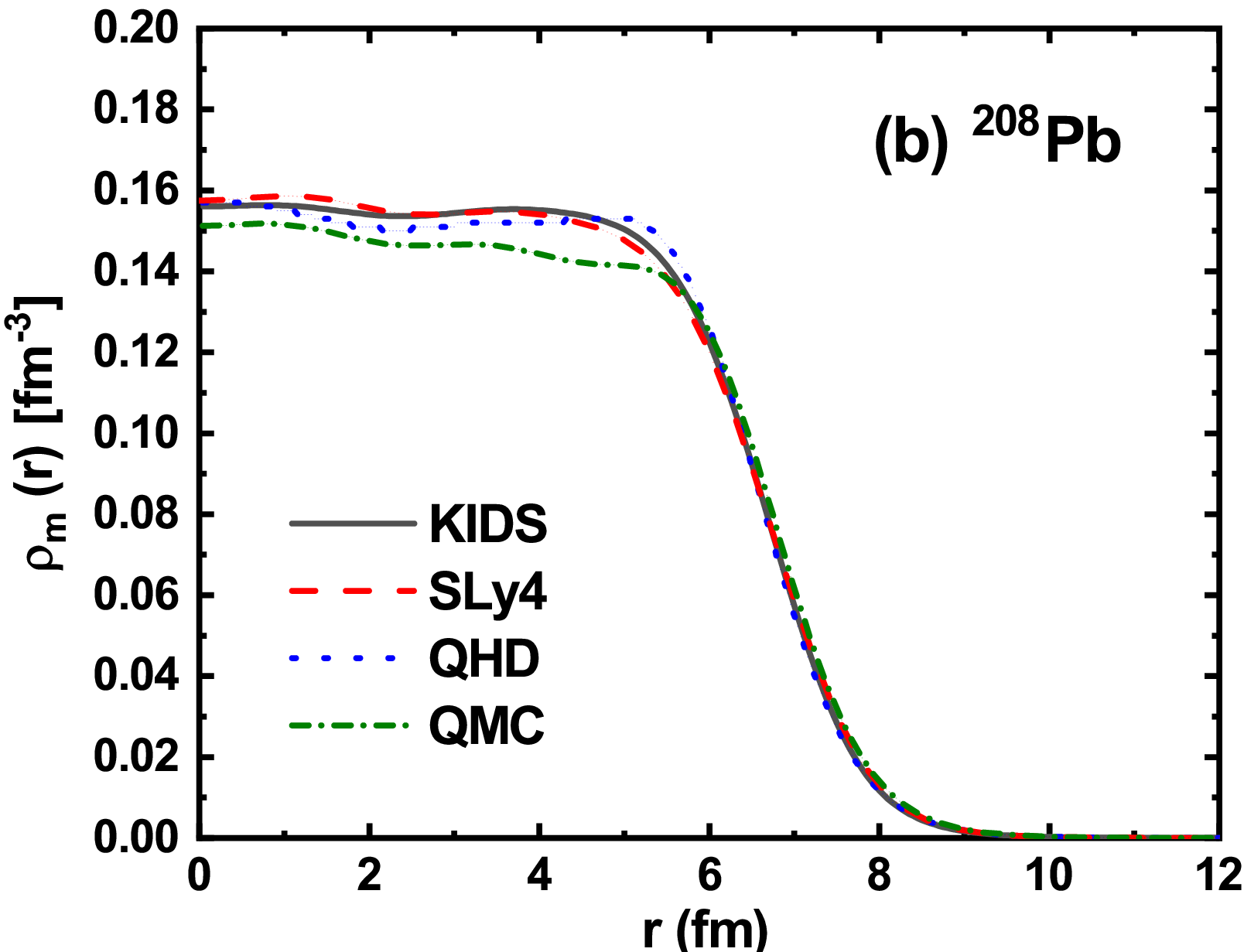}
\includegraphics[width=0.495\linewidth] {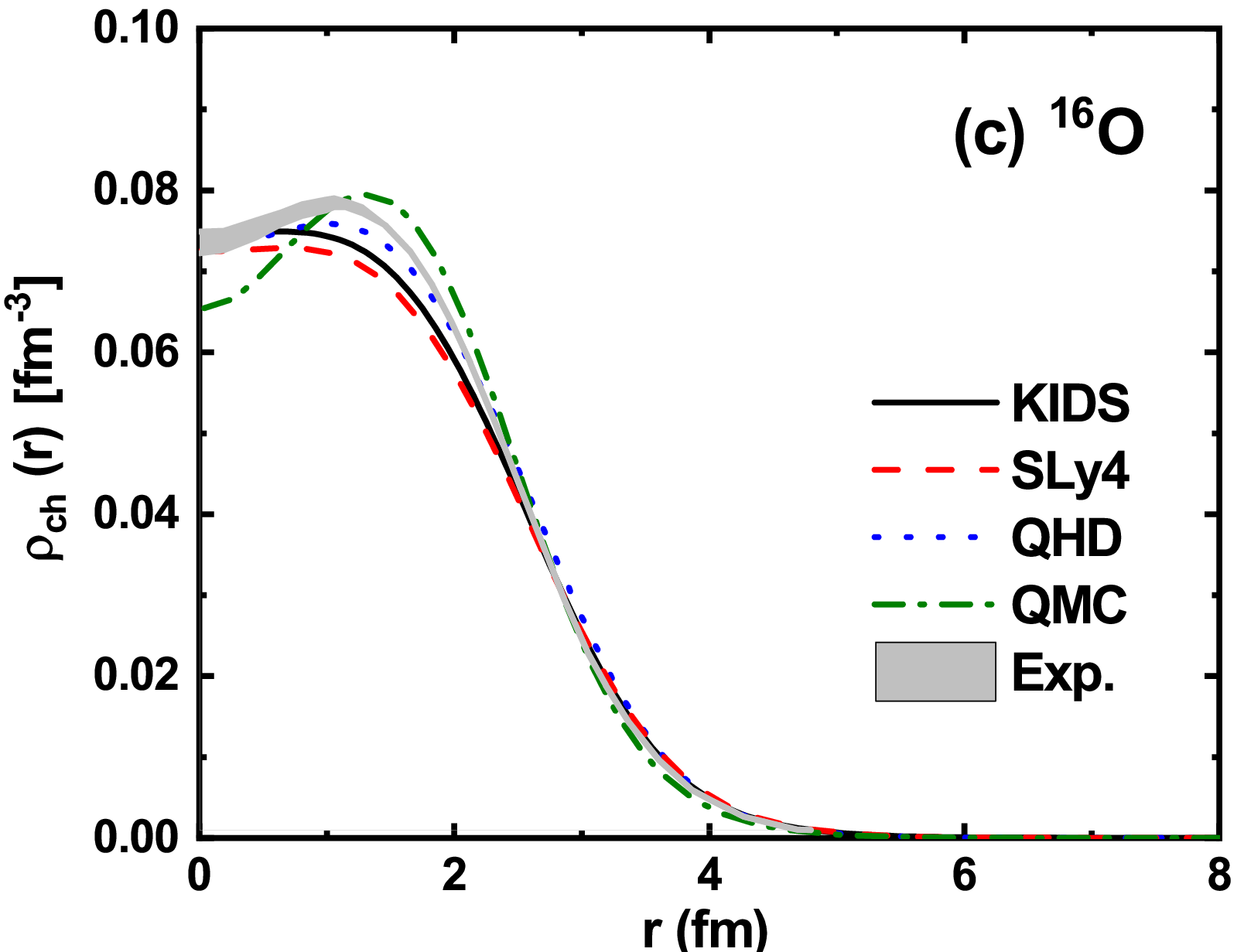}
\includegraphics[width=0.495\linewidth] {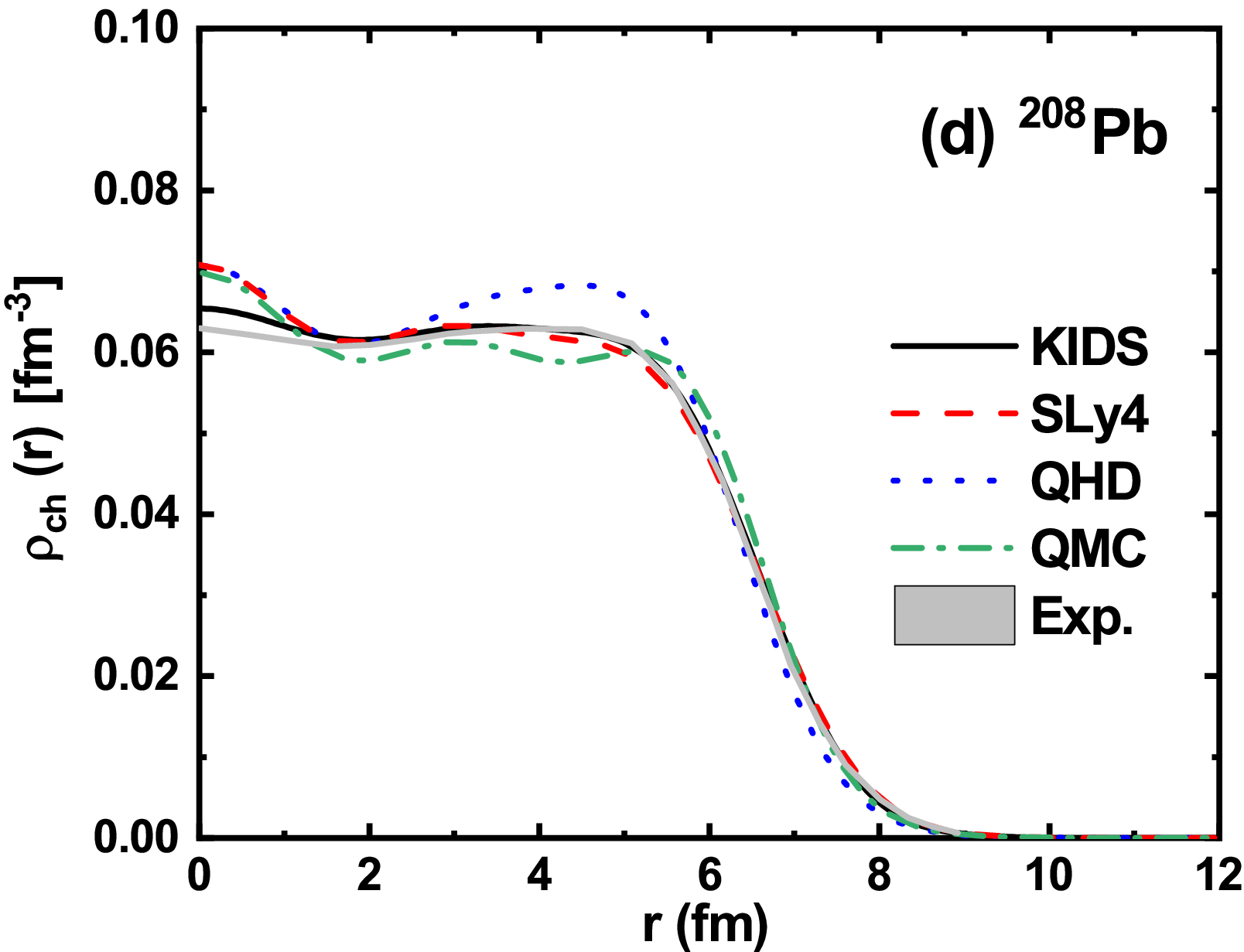}
\end{center}
\caption{\label{fig1}
For $^{16}$O and $^{208}$Pb nuclei, the matter densities ($\rho_{\textrm{m}}$) are shown in the upper panels, and the charge densities ($\rho_{\textrm{ch}}$) in the lower panels.
Experimental data of the charge distributions are taken from Refs.~\cite{sick81,ray79,fria73}.}
\end{figure}
%%%%%%%%%%%%

Figure \ref{fig1} shows the density profiles of $^{16}$O and $^{208}$Pb calculated with the four models.
Upper panels show the matter distribution, and the lower ones show the charge distribution.
Charge distribution is compared with the measurement.
For the matter distribution, in the regions $r \gtrsim 2.5$ fm for $^{16}$O and $r \gtrsim 6$ fm for $^{208}$Pb,
density profiles are similar for all the models, so the model dependence is negligible.
In the inside of these radii, KIDS0 and SLy4 models are close to each other,
but the QHD and the QMC models exhibit distinct behaviors.
QHD model is distinguished from the nonrelativistic models for $^{16}$O, but it is similar with them for $^{208}$Pb.
It will be shown that the difference and similarity have a correlation with the shape of the folding potential.

Charge density of $^{16}$O displays, similar to the matter density, model independence at $r \gtrsim 2.5$ fm,
and the agreement to experimental data is excellent.
At $r \lesssim 2.5$ fm, results are sensitive to the models,
and the QHD model shows a most similar behavior to the measured distribution.
However, considering that $^{16}$O is a light nucleus for which the mean field approximation is less validated than heavy nuclei,
model dependence and deviation from the experimental data in the internal region $r \lesssim 2.5$ fm is reasonable and acceptable.
For the charge density of $^{208}$Pb, in the outer region $r \gtrsim 6$ fm, models are similar to each other.
Model independence is weakened in the internal region $r \lesssim 6$ fm, so the KIDS0, SLy4 and QMC models are similar
and the QHD model behaves separately from the three models.

As a whole, in the outer region that corresponds to surface area, dependence on the model is negligible,
and the agreement to experimental data is accurate.
Internal region shows a stronger dependence on the model.
What effect the dependence will have in the folding potential is examined
in the following subsection.

\subsection{Folding potential}

%%% Figure 2%%%%
\begin{figure}
\begin{center}
\includegraphics[width=0.5\linewidth] {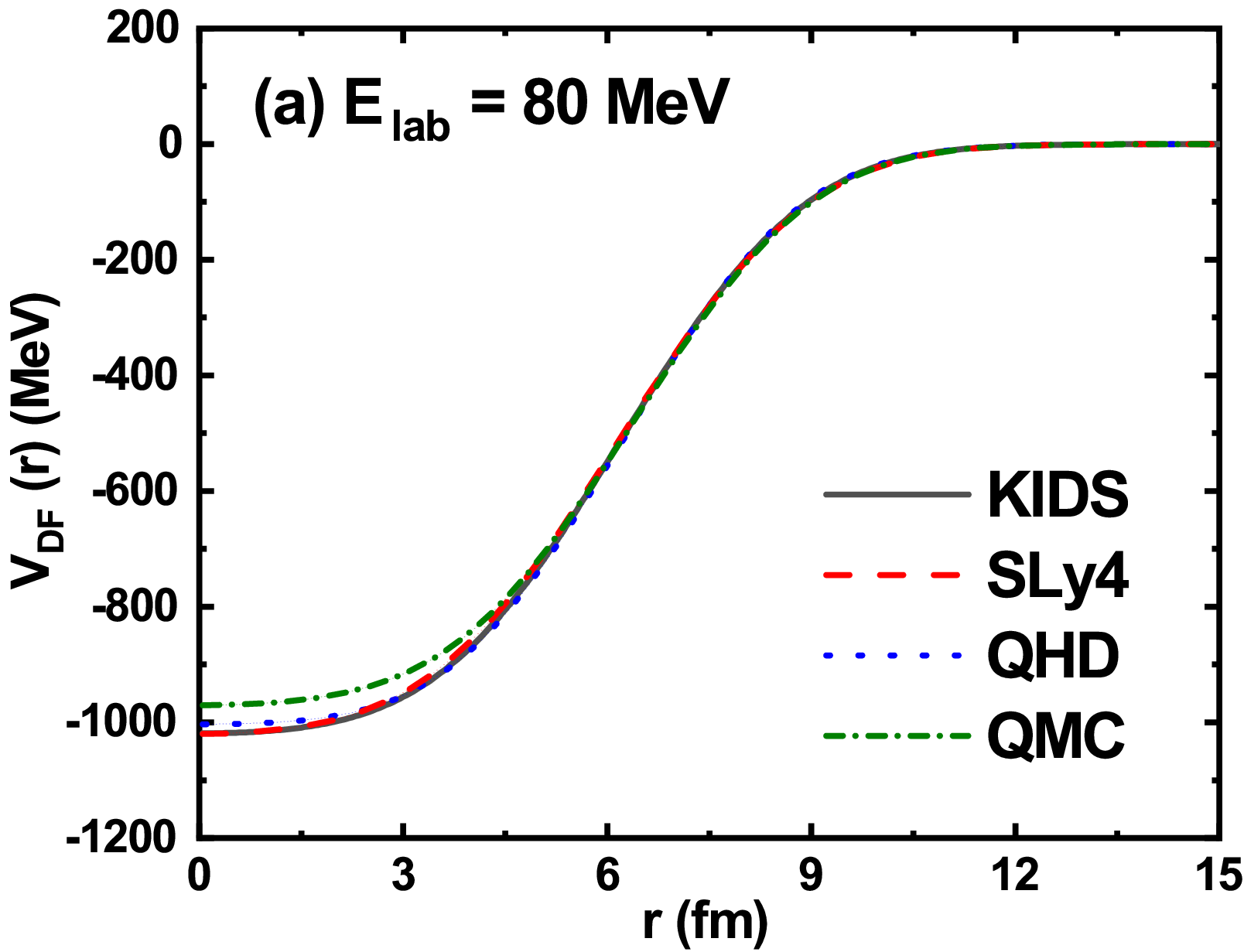}
\includegraphics[width=0.5\linewidth] {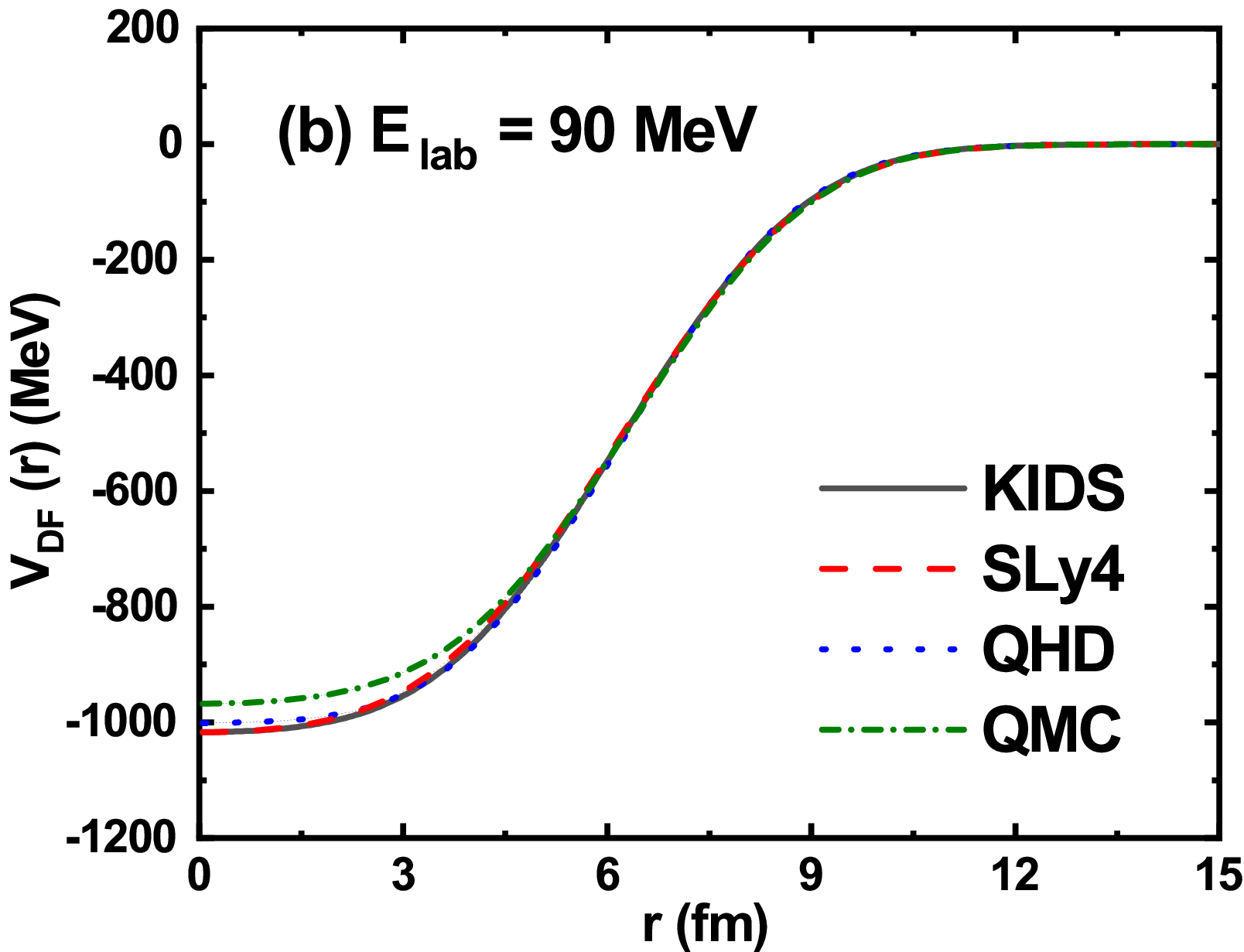}
\includegraphics[width=0.5\linewidth] {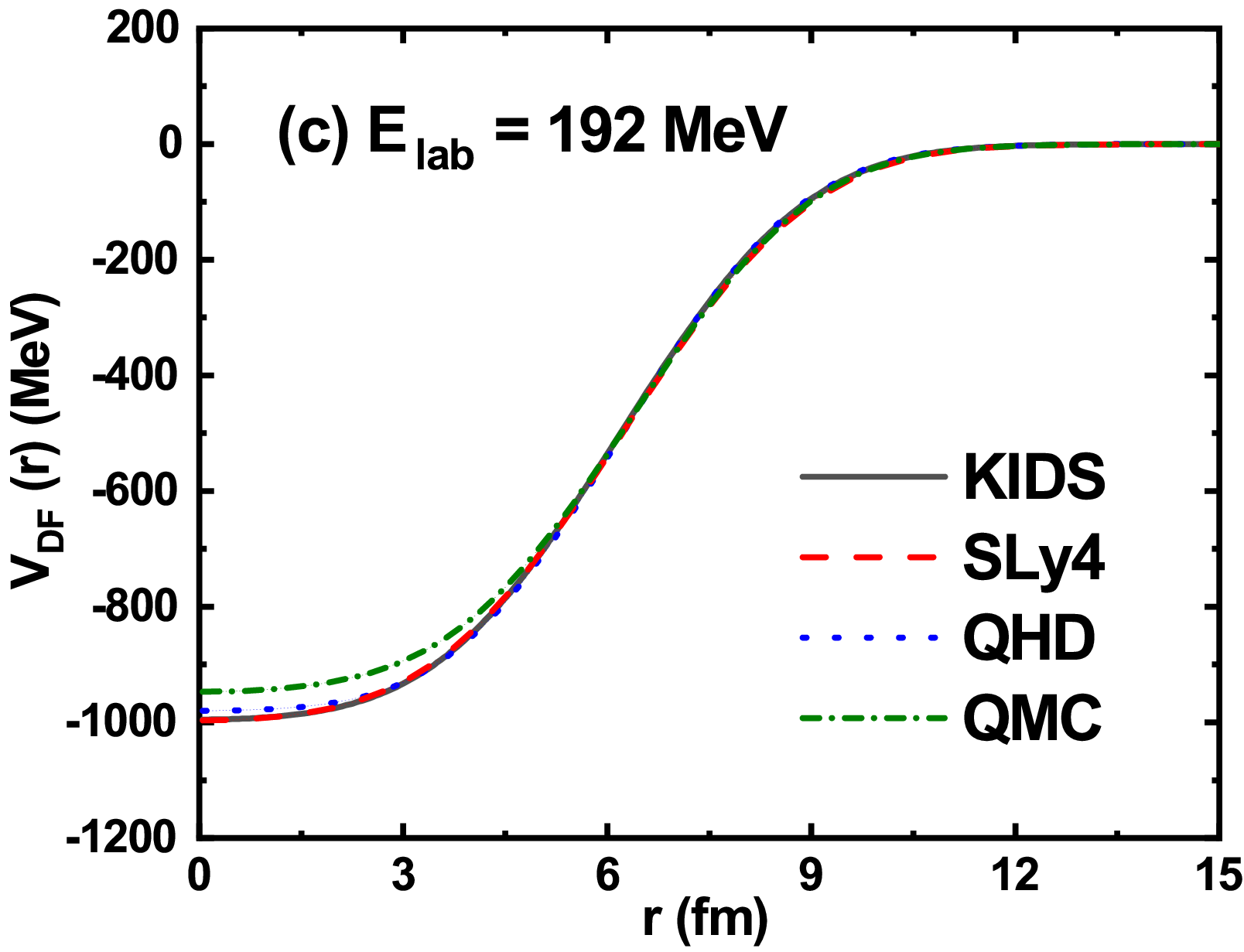}
\end{center}
\caption{\label{fig2}
Double folding potentials $V_{\textrm{DF}} (r)$ for the $^{16}$O+$^{208}$Pb system at the incident energies (a) $E_{\textrm{lab}}$ = 80 MeV (top), (b) $E_{\textrm{lab}}$ = 90 MeV (middle) and (c) $E_{\textrm{lab}}$ = 192 MeV (bottom).}
\end{figure}
%%%%%%%%%%%%

Double folding potentials obtained from the four models are depicted in Fig.~\ref{fig2}
at the incident energies $E_{\textrm{lab}}$ = 80 MeV (top), 90 MeV (middle) and 192 MeV (bottom).
It can be easily seen that the energy dependence is very weak, so the three figures look almost identical.
Looking into more details, the potential value at $r=0$ for $E_{\textrm{lab}}$ = 192 MeV is slightly higher than that for $E_{\textrm{lab}}$ = 80 MeV.
This tiny difference and the weak dependence on the energy could be understood from the M3Y potential.
Energy dependence of the folding potential stems from the delta-function term in Eq. (\ref{v_nn}).
Since the coefficient multiplied to energy is a small number 0.005,
even if $A_1=16$ and $E_{\textrm{lab}}$ = 192 MeV, $1-0.005 E_{\textrm{lab}}/A_1 = 0.94$,
so the correction due to the energy dependence gives a small decrease in magnitude.
Since the delta-function term gives a negative contribution to the folding potential,
decrease in magnitude results in shallower potential depth.
In addition, since $1-0.005 E_{\textrm{lab}}/A_1 = 0.975$ at $E_{\textrm{lab}}$ = 80 MeV, the difference from $E_{\textrm{lab}}$ = 192 MeV is 3.5\%,
so the energy dependence turns out to be very weak.

Comparing the models, KIDS0 and SLy4 overlap completely over the whole region.
QHD begins deviation from KIDS0 and SLy4 at $r \simeq 2$ fm, and the small difference persists to $r=0$.
QMC model shows more sizable difference from the other three models.
The difference for the QMC models starts to appear at around $r\simeq 4.5$ fm.
Since the matter density is used in the calculation of the folding potential,
difference between the models could be understood from Fig.~\ref{fig1} (a) and (b).
For the matter distribution of $^{16}$O, both QHD and QMC deviate from KIDS0 and SLy4 at $r \lesssim 2.5$ fm,
and for $^{208}$Pb, only QMC shows obvious difference at $r \lesssim 6$ fm.
Since the double folding potential is obtained by integrating the densities, difference in the double folding potential becomes visible
when the difference in the density is sufficiently accumulated in the integration.
Therefore, the difference in the double folding potential of the QHD model, though not significant, is shown at $r\lesssim 2$ fm.
For QMC, model dependent region is $r \lesssim 6$ fm, and the difference in double folding potential appears at $r\lesssim 4.5$ fm.

Before proceeding to the reaction process, we note that the double folding potentials obtained from the four models
have a deep depth ($\sim$ 1000 MeV).
For the use in the OM analysis of elastic scattering cross section data~\cite{sat1,keel},
a normalization factor is required to reduce the depth of the double folding potential.
The reduced normalization factor is used in the weakly bound nuclei such as $^{6}$Li~\cite{keel}.

\subsection{Application to the $^{16}$O+$^{208}$Pb reaction}

%%%Table 1%%%
\begin{table}
\begin{center}
\begin{ruledtabular}
\begin{tabular}{cccc}
$E_{\textrm{lab}}$   &   Model       &  $W_{\textrm{I}}$  &   $\chi^{2}$  \\
  (MeV)              &               &        (MeV)       &               \\ \hline
          80         &   SLy4        &        9.705       &      0.207    \\
                     &   QHD         &       11.130       &      0.309    \\ \hline
          90         &   SLy4        &       12.068       &      8.458    \\
                     &   QHD         &       13.025       &      3.574    \\ \hline
         192         &   SLy4        &       16.411       &      2.102    \\
                     &   QHD         &       16.763       &      0.805    \\
\end{tabular}
\end{ruledtabular}
\end{center}
\caption{Optimal parameters of the imaginary potential with the volume-type Woods-Saxon form
in Eq.~(\ref{om_pot}) for the elastic cross sections of $^{16}$O+$^{208}$Pb system obtained from the SLy4 and QHD models.
Units of $E_{\rm lab}$ and $W_{\rm I}$ are MeV. Note that in order to obtain the depth of the imaginary potential, we fix $a_{\rm I}=0.600$ fm and $r_{\rm I}=1.300$ fm.}
\label{parameters_fold}
\end{table}

Using the double folding potential obtained from the four models, we calculate the elastic scattering cross sections of $^{16}$O+$^{208}$Pb
at the laboratory energies $E_{\textrm{lab}}$ = 80, 90 and 192 MeV.
There are several methods to solve the core reaction.
In our study, we analyze it by using a simple optical model~\cite{so3}.
We solve the Schr\"{o}dinger equation
\begin{equation} \label{scodinger}
[E - T_{l} (r)] \chi^{(+)}_{l} (r) = U_{\textrm{OM}} (r)~\chi^{(+)}_{l} (r)
\end{equation}
with the OM potential given by
\begin{eqnarray} \label{om_pot}
U_{\textrm{OM}} (r) &=& V_{\textrm{C}} (r) + NV_{\textrm{DF}} (r) - i W(r) \nonumber \\
                    &=& V_{\textrm{C}} (r) + NV_{\textrm{DF}} (r) - \frac{i W_{\textrm{I}}}{1+\exp(\frac{r-R_{\textrm{I}}}{a_{\textrm{I}}})}.
\end{eqnarray}
Here $T_{l} (r)$ and $\chi^{(+)}_{l} (r)$ are a kinetic energy operator and a distorted partial wave function
with the angular momentum $l$, respectively.
The OM potential $U_{\textrm{OM}} (r)$ in Eq.~(\ref{om_pot}) is composed of a Coulomb potential $V_{\textrm{C}} (r)$,
a double folding potential $V_{\textrm{DF}} (r)$, and an imaginary Woods-Saxon potential with a volume-type $W(r)$.
Note that we will take the normalization factor $N$ = 1 to simplify the analysis of the elastic scattering data.

\begin{figure}
\begin{center}
\textbf{SLy4}  \hspace{8cm} \textbf{QHD} \\
\includegraphics[width=0.495\linewidth] {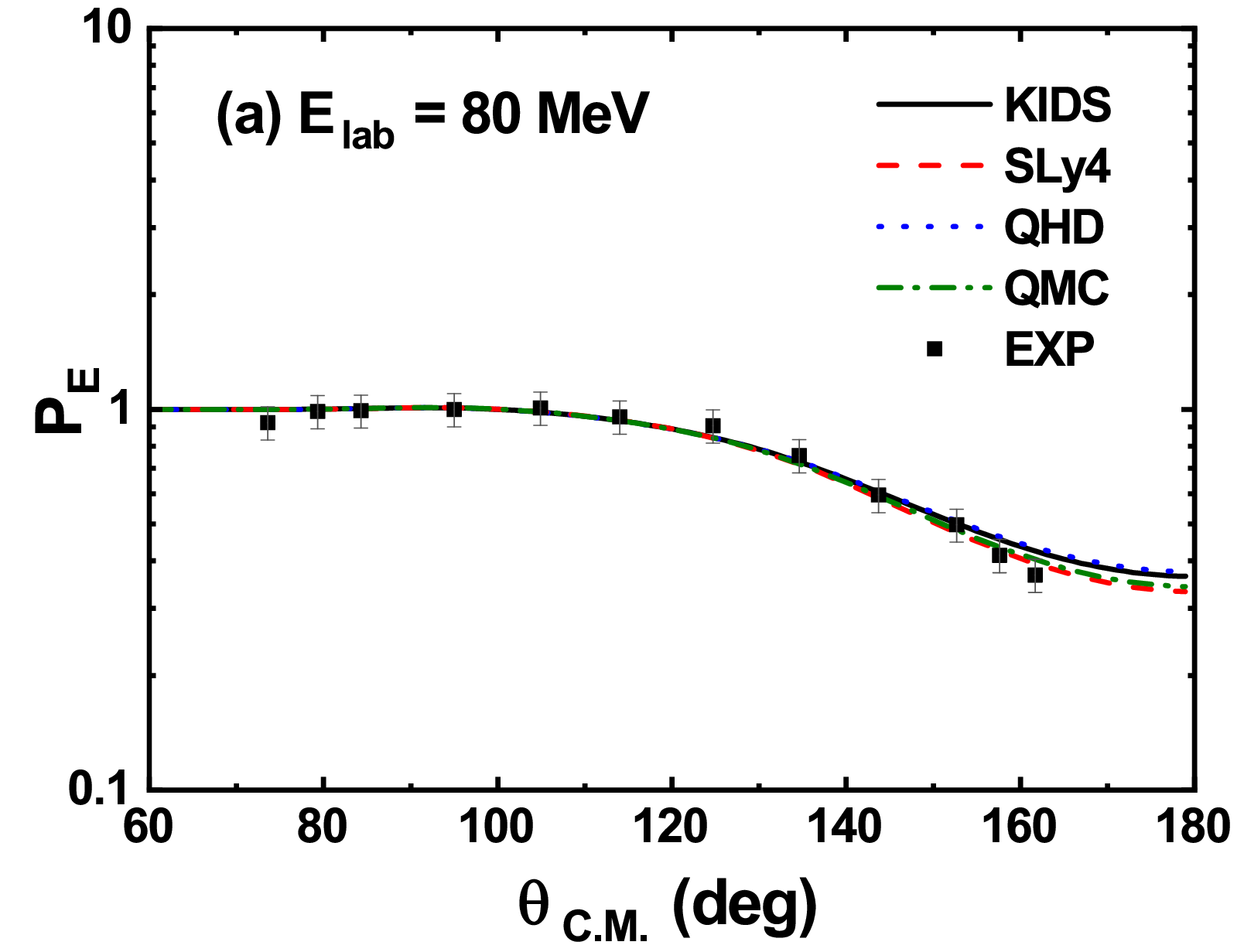}
\includegraphics[width=0.495\linewidth] {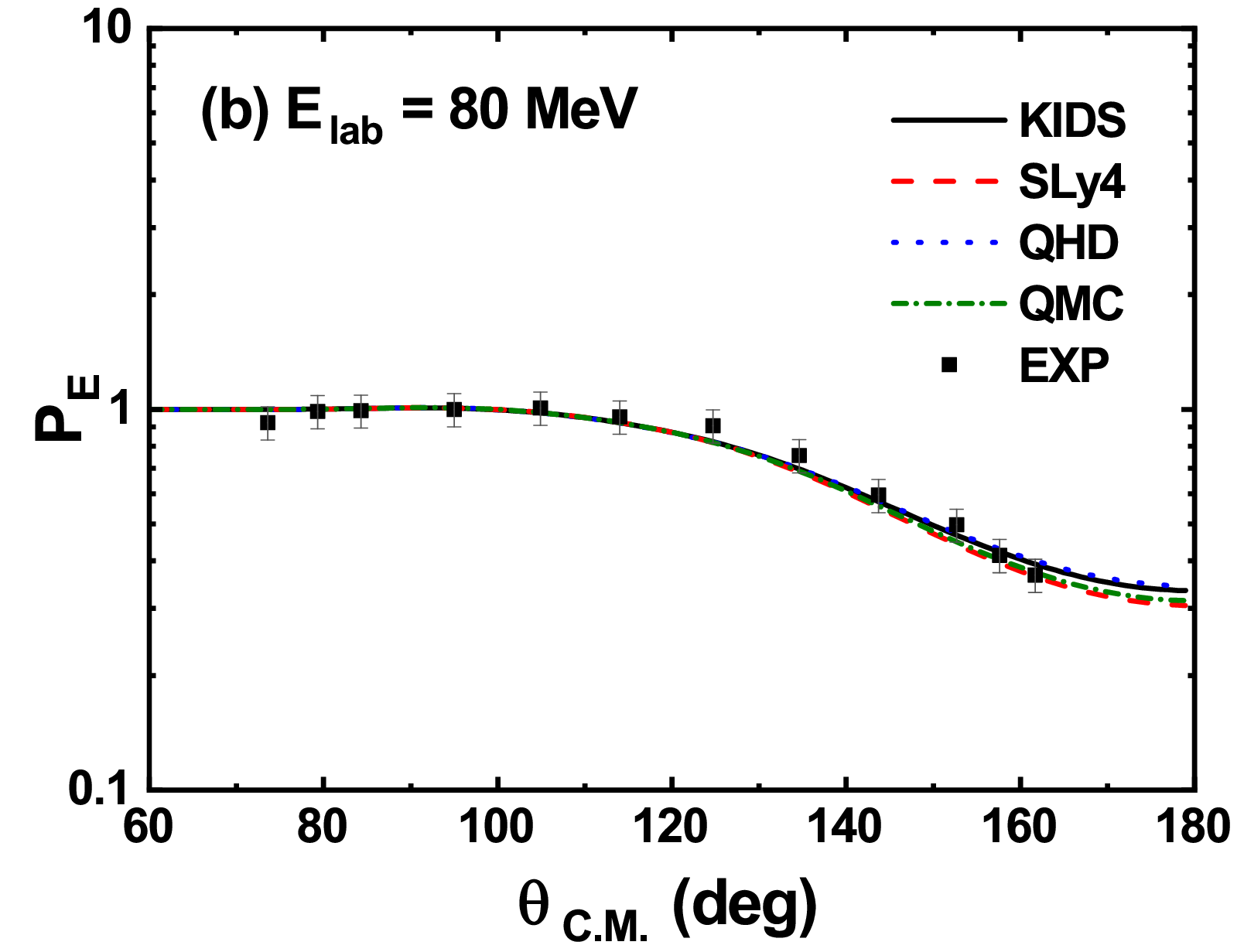}
\includegraphics[width=0.495\linewidth] {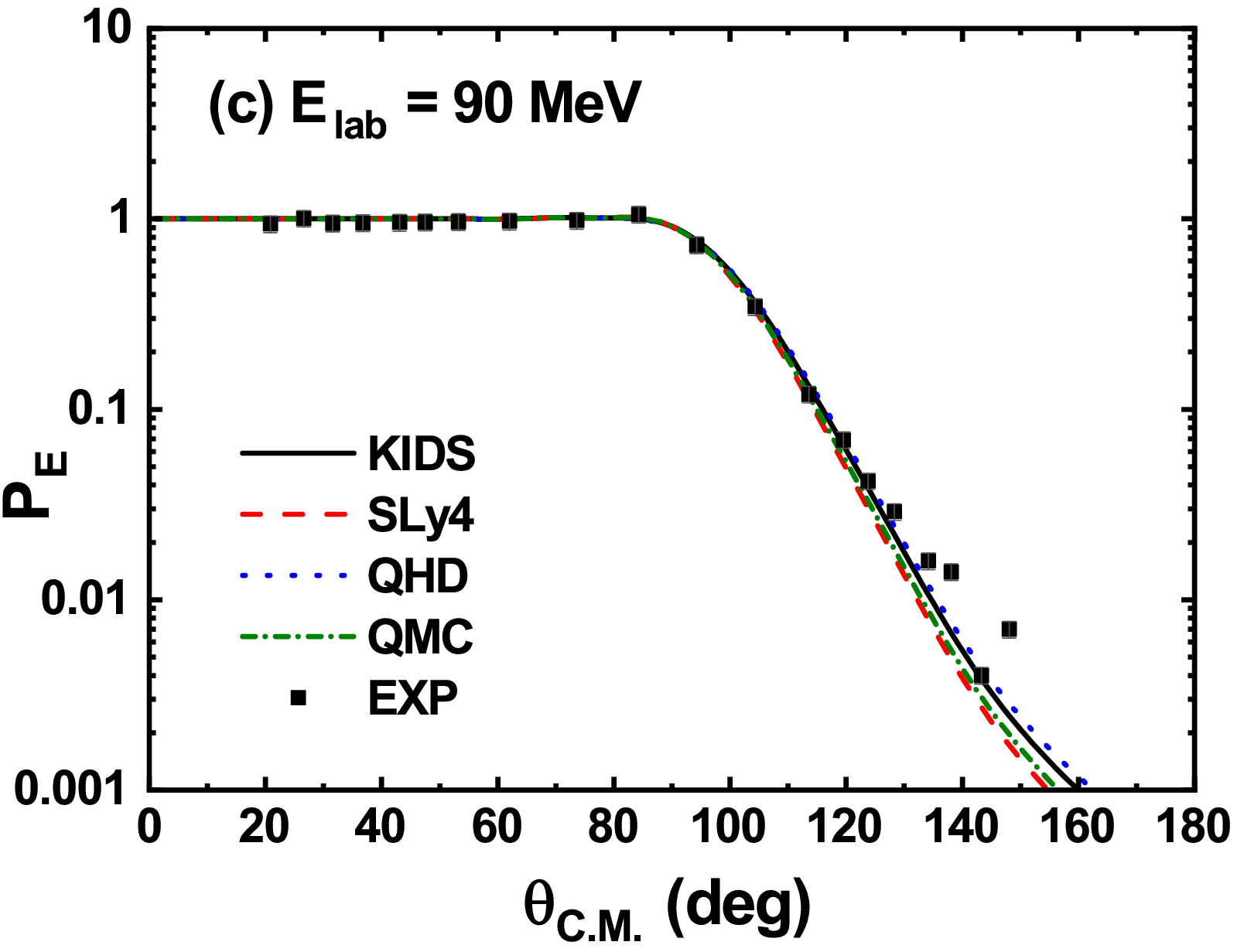}
\includegraphics[width=0.495\linewidth] {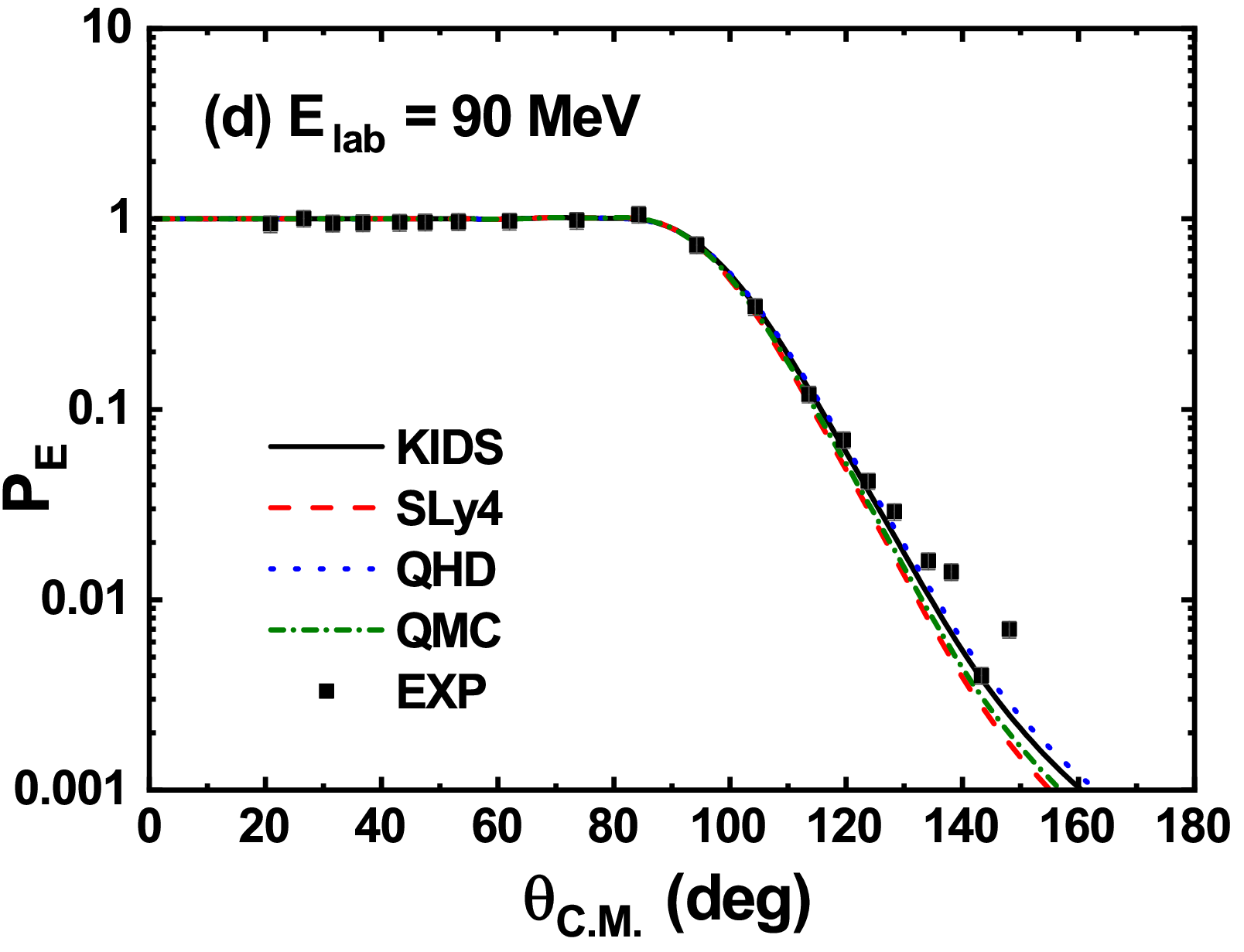}
\includegraphics[width=0.495\linewidth] {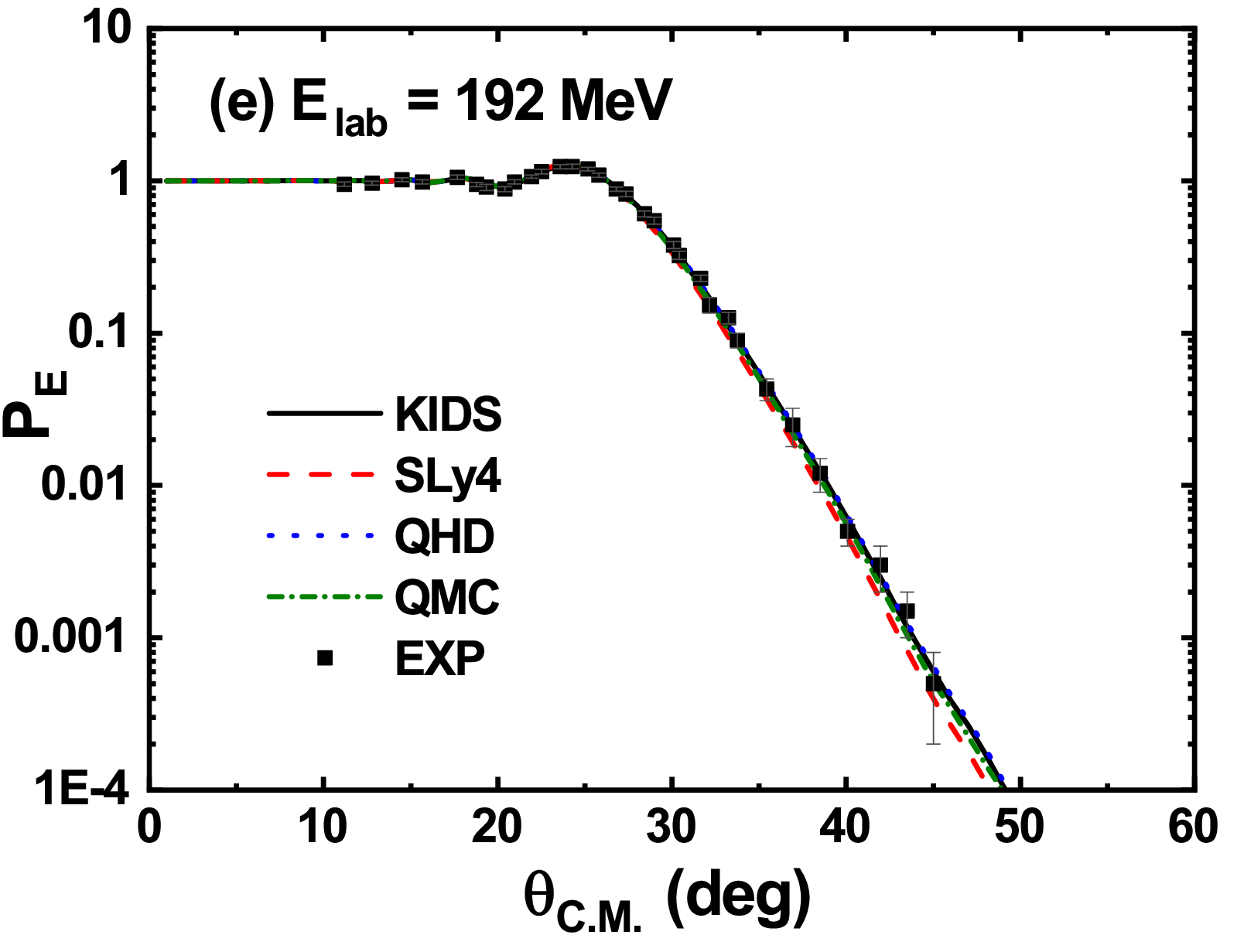}
\includegraphics[width=0.495\linewidth] {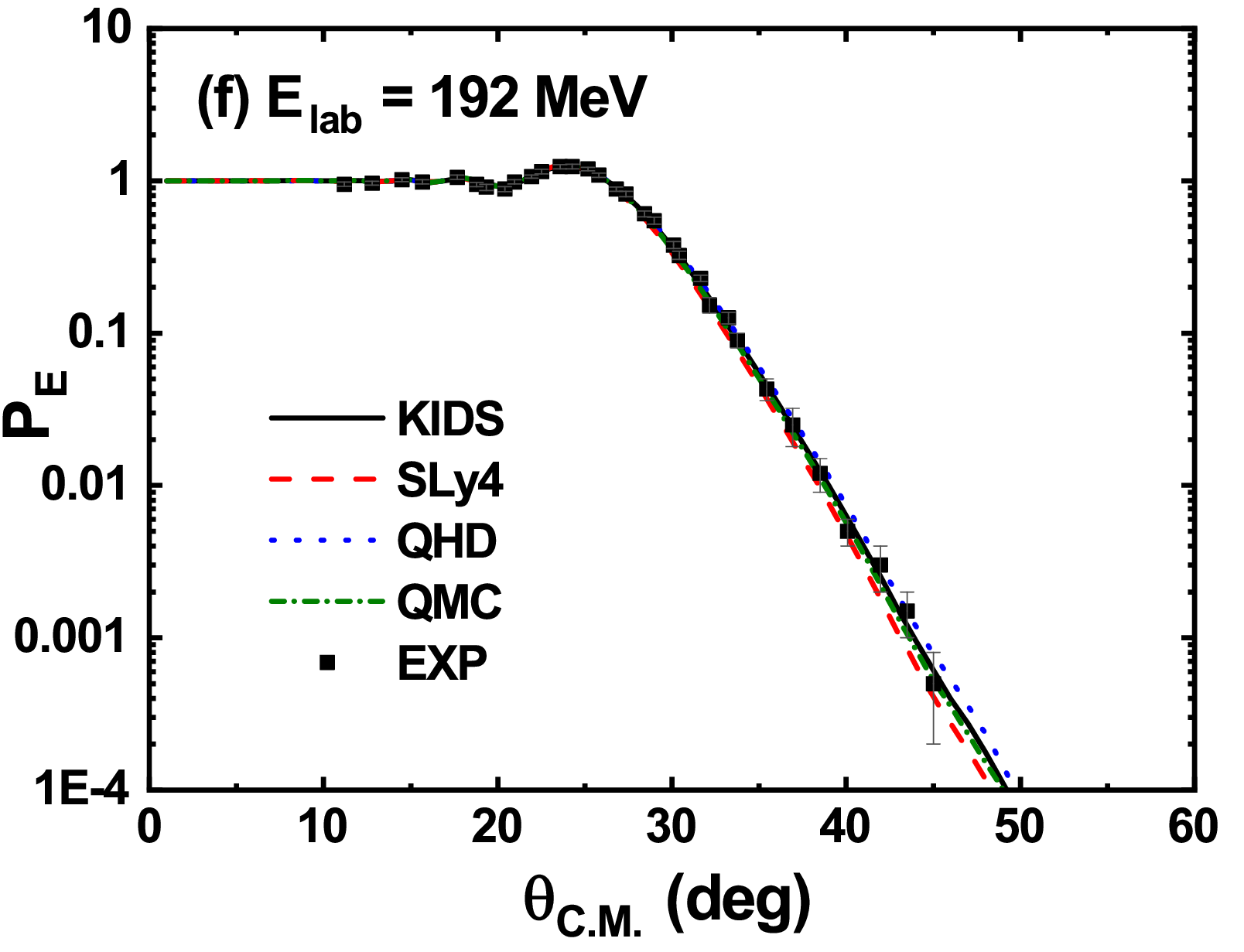}
\end{center}
\caption{\label{fig3}
The ratios of the elastic scattering cross section
to the Rutherford cross section for the $^{16}$O+$^{208}$Pb system at the incident energies $E_{\textrm{lab}}$ = 80 MeV, 90 MeV and 192 MeV
are shown in the top, middle and bottom panels, respectively.
Panels in the left columns show the result with the imaginary potential fitted with the SLy4 model,
and the panels in the right column correspond to the result with the imaginary potential fitted with the QHD model.
Experimental elastic scattering cross section data are taken from Refs.~\cite{vide77,vulg86,ball75}}.
\end{figure}

In order to investigate the analysis of the elastic scattering cross section data for $^{16}$O+$^{208}$Pb system at
$E_{\textrm{lab}}$ = 80\;MeV, 90\;MeV, and 192\;MeV,
we perform the least $\chi^2$-fitting of the imaginary potential to the experimental elastic scattering cross section data.
In order to check the dependence on the model, we use SLy4 and QHD models to fit the parameters in the imaginary potential.
To reduce the number of parameters, we fix $a_{\rm I}=0.600$ fm and $r_{\rm I}=1.300$ fm
($r_{\rm I} = R_{\rm I}/(A^{1/3}_1 + A^{1/3}_2)$ where $A_1$ and $A_2$ are the mass number of projectile and target nuclei, respectively)
and determine $W_{\rm I}$ from the fitting.
The extracted parameter sets are tabulated in Tab.~\ref{parameters_fold}.
It is notable that since the double folding potentials obtained from the four models are similar,
the parameter sets of the imaginary potential extracted through least $\chi^2$-fitting are also similar at a given energy.
The similarity implies that the imaginary potential obtained from one model can be applied to other models as an approximation.
To check the validity of this approximation, we apply the imaginary potentials determined with the SLy4 (or QHD) model
to the KIDS0, QMC, and QHD (or KIDS0, QMC, and SLy4) models in the calculation of the cross section.
$\chi^2$ values at $E_{\textrm{lab}}$ = 90 MeV are larger than the other energies by an order of magnitude.
The reason could be understood from Fig.~\ref{fig3}.
Three data between 130 and 150 degrees for $E_{\textrm{lab}}$ = 90 MeV show a bumping behavior,
so they cannot be easily reproduced with a simple Woods-Saxon form of the imaginary potential.
Inconsistency between experiment and theory for the three data causes large $\chi^2$ value for $E_{\textrm{lab}}$ = 90 MeV.

%%%Figure 3%%%

Using the optimal parameters in Tab.~\ref{parameters_fold}, next, we calculate the ratio of
the elastic scattering cross section ($\sigma_{\textrm{E}}$) to the Rutherford one ($\sigma_{\textrm{RU}}$), $P_{\textrm{E}}=\sigma_{\textrm{E}}/\sigma_{\textrm{RU}}$,
as a function of the center-of-mass angle ($\theta_{\textrm{c.m.}}$) for the $^{16}$O+$^{208}$Pb system at
$E_{\textrm{lab}}$ = 80\;MeV, 90\;MeV, and 192\;MeV
and compare the theoretical $P_{\textrm{E}}$ values with the experimental ones.

Figure \ref{fig3} displays the result.
Panels in the left column are the result with $W_{\rm I}$ fitted with the SLy4 model,
and the panels in the right column are the result with $W_{\rm I}$ fitted with the QHD model.
Panels in the top, middle and bottom rows correspond to $E_{\rm lab}=$ 80, 90 and 192 MeV, respectively.
A common feature is that the results are essentially independent of the model in the forward angles,
and the model dependence becomes visible at substantially backward angles.
However, considering that the vertical axes are in the log scale,
the difference between the models in the backward angle is not significant.
This is a most important finding of the work: within the EDF models considered in the work,
theoretical predictions are not sensitive to a specific model.
Even if an imaginary potential obtained from a model is applied to the other model,
though it is not a fully consistent calculation, the result can maintain an accuracy similar to the original model.

From the results in Fig.~\ref{fig3}, the difference of $W_I$ between the SLy4 and QHD models decrease with higher incident energies.
Model dependence in the cross section also decreases as energy increases,
so one may deduce that the difference and similarity of $W_{\rm I}$ could be a prime source of the difference in the result.
For $E_{\rm lab}=$80 and 90 MeV, at the angles where the model dependence is evident,
there is a tendency that KIDS0 and QHD are similar, and SLy4 and QMC form another group of similar behavior.
This grouping of the models is not directly correlated to the matter distribution because KIDS0 model is most similar to SLy4 model
for both $^{16}$O and $^{208}$Pb, but they do not show such a high similarity in the elastic scattering cross section.

It may be hasty and premature to draw a definite and general conclusion from the results in the work.
However, considering that the four models are very different in their basic formalisms, in the ways to determine the model parameters,
and in obtaining wave functions (i.e. Hartree-Fock vs. Dirac),
weak dependence on the model could be a global feature of the nuclear density functional approach.
Of course, such a model insensitivity will face limitations at some point.
Those limitations could be figured out by applying the model to various reaction processes and diverse kinematic conditions.

\section{Summary and Conclusion}
The double folding potentials are constructed using the M3Y interaction and the matter densities obtained from the four EDF models.
The elastic scattering cross sections are calculated using the obtained double folding potentials through the OM.
The effect of difference in (or dependency on) the matter densities obtained from the four models to the elastic scattering cross sections is then investigated.

For this purpose, we calculated the matter and charge densities using four EDF models.
The results indicate that the density distributions differ slightly in the internal region ($r$ $\lesssim$ 2.5 fm for $^{16}$O and $r$ $\lesssim$ 6 fm for $^{208}$Pb),
depending on the model. However, the difference of the density distributions is negligible in the outer region
($r$ $\gtrsim$ 2.5 fm for $^{16}$O and $r$ $\gtrsim$ 6 fm for $^{208}$Pb) corresponding to surface area.

Next, we utilize the obtained matter densities and the M3Y interaction to calculate double folding potentials.
The double folding potentials exhibit a slight variation in the depth of the potential in the internal region ($r$ $\lesssim$ 6 fm).
Notably, the double folding potential derived from the QMC model differs significantly from the other three ones.
However, we observe the insensitivity to the nuclear model in the outer regions ($r$ $\gtrsim$ 6 fm).
Additionally, the alteration in the double folding potential with incident energy is imperceptible.

Finally, the double folding potentials from the four EDF models are applied to the OM.
The elastic scattering cross section generated by using the OM for these double folding potentials shows no model dependency,
except for a slight difference in the backward angle (or the internal region).
This result is originated from that the matter densities obtained from the four EDF models produce nearly identical imaginary Woods-Saxon potentials in the outer region (or the forward angle),
where the elastic scattering cross section is dominant.
Consequently, the elastic scattering cross section at the forward angle exhibits minimal model dependency.

\section*{ACKNOWLEDGMENTS}
This work was supported by the National Research Foundation of Korea
(Grant Nos. NRF-2021R1F1A1046575, 2021R1F1A1051935, 2023R1A2C1003177, RS-2023-00276361) and by MSIT (No. 2018R1A5A1025563).
We thank Dr. Soonchul Choi for providing the wave functions of the QMC model.

\end{document}